# Prediction of Autism Treatment Response from Baseline fMRI using Random Forests and Tree Bagging

Nicha C. Dvornek[1], Daniel Yang[1], Archana Venkataraman[3], Pamela Ventola[1], Lawrence H. Staib[2,4,5], Kevin A. Pelphrey[6], James S. Duncan[2,4,5]

[1]Child Study Center and [2]Department of Radiology & Biomedical Imaging, Yale School of Medicine, New Haven, CT, USA
[3]Department of Electrical and Computer Engineering, Johns Hopkins University, Baltimore, Maryland, USA
[4]Department of Biomedical Engineering and [5]Department of Electrical Engineering, Yale University, New Haven, CT, USA
[6]Autism and Neurodevelopmental Disorders Institute, George Washington University and Children's National Medical Center, Washington, DC, USA

**Abstract.** Treating children with autism spectrum disorders (ASD) with behavioral interventions, such as Pivotal Response Treatment (PRT), has shown promise in recent studies. However, deciding which therapy is best for a given patient is largely by trial and error, and choosing an ineffective intervention results in loss of valuable treatment time. We propose predicting patient response to PRT from baseline task-based fMRI by the novel application of a random forest and tree bagging strategy. Our proposed learning pipeline uses random forest regression to determine candidate brain voxels that may be informative in predicting treatment response. The candidate voxels are then tested stepwise for inclusion in a bagged tree ensemble. After the predictive model is constructed, bias correction is performed to further increase prediction accuracy. Using data from 19 ASD children who underwent a 16 week trial of PRT and a leave-one-out cross-validation framework, the presented learning pipeline was tested against several standard methods and variations of the pipeline and resulted in the highest prediction accuracy.

## 1 Introduction

Autism spectrum disorders (ASD) are a group of neurological developmental disorders that are characterized by impaired social interactions, difficulties in communication, and repetitive behaviors [1]. Current treatments include behavioral-based therapies aimed at treating these core symptoms of autism. One such therapy is Pivotal Response Treatment (PRT), an empirically-supported practice that targets pivotal areas of development, such as motivation and self-initiation [11]. Behavioral therapies require large time commitments, including patient sessions with therapists, training for care providers, and lifestyle changes for the

This work was supported in part by T32 MH18268 and R01 NS035193.

families of ASD children. These intensive interventions have shown promise in recent studies [6,17]. However, given the complexity of ASD, an intervention that works well for one patient may not be effective in another, and choosing the right therapy is largely by trial and error. Given the importance of early intervention and the commitment behavioral interventions require, the ability to predict a treatment's effectiveness for a given patient would be extremely valuable.

Functional magnetic resonance imaging (fMRI) has helped characterize neural networks and brain changes that occur in ASD [9,18,10]. As fMRI has aided the understanding of ASD pathophysiology, we propose using task-based fMRI to predict response to ASD therapy. While fMRI has been employed for predicting changes in autistic traits [13] and treatment outcomes in other brain disorders [2], it has not yet been used for predicting ASD treatment response.

A challenge of training a model to predict autism treatment response from fMRI data is the large number of possible inputs, combined with the small number of subjects due to difficulties imaging ASD children. A popular machine learning technique that has emerged for such "large $p$, small $n$" problems is random forests [15]. The algorithm's use of random feature subset selection makes it suitable for exploring high dimensional inputs while improving prediction accuracy by reducing the variance of the estimate. However, including noisy inputs into the random forest still degrades prediction accuracy. In addition, the small sample sizes in autism studies contribute to low prediction strength of the individual trees, and randomized feature selection produces even weaker trees.

In this work, we describe the novel application of random forest regression and bagged tree ensembles for predicting an ASD patient's response to therapy from baseline fMRI. The proposed learning pipeline aims to address the above described difficulties by utilizing random forests for candidate variable selection, further refining the input variables stepwise to build a bagged tree ensemble, and performing bias correction to improve predictions at the extremes of the outcome range. We tested the proposed algorithm against standard methods and variations of the pipeline on data from ASD children treated with PRT using a leave-one-out cross-validation approach to assess prediction accuracy.

## 2 Methods

### 2.1 Input and Output Definitions

The input variables for the learning pipeline are derived from fMRI scans acquired prior to PRT. We utilized a motion perception task-based fMRI paradigm that has highlighted activation and functional connectivity differences in ASD children, their unaffected siblings and typical controls [10,16]. During the scans, subjects viewed point light animations of coherent and scrambled biological motion in a block design [10]. The t-statistic images were computed for the contrast of biological motion greater than scrambled motion and mapped to MNI152 space [7]. Because PRT aims to improve core deficits in social motivation, we focus the analysis on brain regions associated with social motivation: the or-

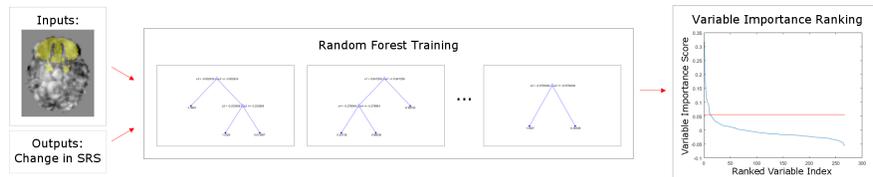

(a) Candidate Variable Selection using Random Forests

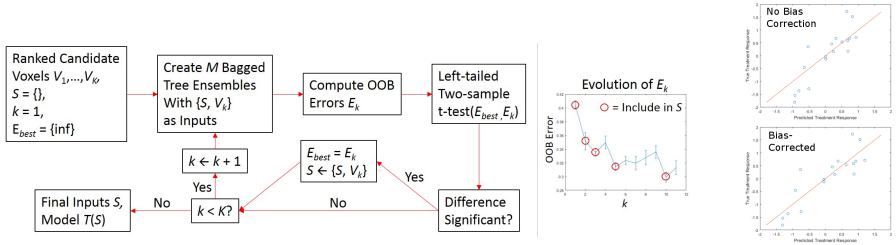

(b) Stepwise Building of Bagged Tree Ensemble    (c) Bias Correction

Fig. 1: Overview of the proposed learning pipeline.

bitofrontal cortex, ventromedial prefrontal cortex, amygdala, and ventral striatum [4]. The t-statistics for voxels from these regions were used as inputs.

The output of the predictive model is the patient response to PRT, based on the Social Responsiveness Scale, Second Edition (SRS) score [5]. The SRS measures the severity of social impairment associated with ASD, with lower scores indicating better social function. The change in SRS score measured before and after PRT, controlled for baseline severity, is set as the regression target.

### 2.2 Learning Pipeline

Fig. 1 gives an overview of our proposed learning pipeline. First, random forests rank voxel importance, and voxels that may be good predictors are selected. The candidate voxels are then tested stepwise in order of importance for inclusion in a bagged tree ensemble. Finally, bias correction is estimated. Details are below.

**Candidate Variable Selection Using Random Forests** Random forests is an ensemble learning method which combines bagging decision trees with random subset sampling of the predictors for constructing each node split [3]. In random forest regression, the predictions from the individual regression trees are averaged. The error rate of a random forest can be estimated by the out-of-bag (OOB) error. Predicted outcomes for OOB samples for each tree are averaged across trees for which the observation is OOB. The OOB error is then the mean squared error between the OOB ensemble predictions and the true outcomes.

Furthermore, the random forest algorithm can rank variable importance by using the OOB samples. The importance of a variable is measured by calculating the average change in the prediction error for OOB samples for each tree when the values of that variable are randomly permuted across the OOB observations. The greater the increase in error, the higher the importance of the variable. Note that small negative importance scores are possible, in which the permuted values for an irrelevant variable result in a small decrease in error by chance.

We use random forests to obtain a variable importance ranking of all input voxels. We retain those voxels whose importance score is greater than the absolute value of the lowest negative importance score (Fig. 1a). This threshold is chosen for excluding voxels because irrelevant variables will have low importance scores that fluctuate randomly around zero. To ensure a stable importance ranking, we run the random forest algorithm with a large number of trees.

**Stepwise Building of Bagged Tree Ensemble** A standard random forest learning pipeline would construct another random forest using the retained top-ranked variables as input. However, this results in poor predictive power for the ASD treatment data (see results and discussion below). Here, we propose that using bagged trees and refining the set of best input variables rather than using random subset sampling results in more accurate predictions.

To determine which of the $K$ candidate voxels should be kept as inputs for the bagged tree ensemble, we use a stepwise approach, assessing the addition of each variable $V_1, \ldots, V_K$ in order of their importance ranking (Fig. 1b, left). Let $S$ be the set containing the current best voxel inputs after investigating the first $k-1$ voxels. Using the voxels in $S$ as inputs, we created $M$ bagged tree ensembles and calculated the associated OOB errors $E_{best}$. To determine whether the $k$th-ranked variable $V_k$ should be included in the final model, $M$ bagged tree ensembles are trained using the voxels in $S$ and the new feature $V_k$ as inputs. The OOB errors $E_k$ are calculated for the $M$ new ensembles. If the new errors $E_k$ are significantly lower than the errors from the current best model $E_{best}$ according to a one-tailed two-sample t-test, the new voxel $V_k$ is added to the set of best inputs $S$ (Fig. 1b, right). This process is iterated for all $K$ candidate voxels, resulting in the final bagged tree ensemble $T(S)$. In the following, we set $M=10$ and the significance level for the t-tests to 0.05. Then, for new patient data, the selected best voxels in $S$ are extracted and input into the learned model $T(S)$ to produce the predicted treatment response $Y_{ens}$.

**Bias Correction** Tree-based ensembles for regression tend to underestimate the high values and overestimate the low values in the outcome range. This is because the predicted regressor targets are averages of the response values in the training set. Furthermore, because neuroimaging studies of ASD children tend to have small sample sizes, the prediction strength of the trees is weaker, and bias correction may have a significant impact.

To attenuate the bias, we assume a linear relationship between the predicted and true outcome $Y_{true} = \beta_1 Y_{ens} + \beta_0$. Linear regression parameters $\hat{\beta}_1$ and $\hat{\beta}_0$ are estimated using the ensemble predictions for the OOB samples and the true response values (Fig. 1c). The bias-corrected prediction is then $\hat{Y} = \hat{\beta}_1 Y_{ens} + \hat{\beta}_0$.

## 3 Experiments

### 3.1 Data and Image Preprocessing

Data was collected from 19 children (age $M=5.87$ years, $SD=1.09$ years) with ASD (IQ $M=105$, $SD=16.8$) who underwent 16 weeks of PRT. SRS scores were measured at baseline ($M=82.7$, $SD=22.6$) and post-treatment ($M=66.5$, $SD=23.5$). Imaging at baseline included a T1-weighted MP-RAGE structural MRI (isotropic 1 mm$^3$ resolution) and BOLD T2*-weighted fMRI acquired during the point light animation paradigm (164 volumes, voxel size $3.44 \times 3.44 \times 4.00$ mm$^3$). Note data collection involved over 2200 hours of treatment and imaging.

Images were processed in FSL [8]. First, a standardized fMRI preprocessing stream as described by Pruim et al. [14] was applied. Functional images were aligned to the structural image, which was registered to the MNI brain. Next, t-statistic images for the contrast of biological motion greater than scrambled motion were computed and warped to MNI space using the estimated deformations. To reduce computation, images were downsampled by a factor of four. The orbitofrontal cortex, ventromedial prefrontal cortex and amygdala were defined using the Harvard-Oxford structural atlases, and the ventral striatum was defined using the Oxford-GSK-Imanova structural striatal atlas, all in FSL.

### 3.2 Experimental Methods

We compared the proposed learning pipeline of random forest variable selection, stepwise variable refinement for building the bagged tree ensemble, and bias correction (RF-BS-BC) with the following methods:

1. Standard random forest (RF).
2. Standard support vector machine with a linear kernel (SVM), which outperformed tested nonlinear kernels.
3. Variable selection using random forests as in Section 2.2, followed by random forest training using the selected variables as inputs (RF-RF).
4. Variable selection using random forests as in Section 2.2, followed by bagged tree ensemble training using the selected variables as inputs (RF-B).
5. The proposed candidate variable selection and stepwise variable refinement to create the bagged tree ensemble, without bias correction (RF-BS).

Methods were implemented in MATLAB [12]. Random forests were trained using parameter defaults for the number of variables to sample (1/3 the number of inputs). Random forests used for variable selection were run with 5000 trees. Ensembles using selected variables as inputs were created with 1000 trees.

Leave-one-out cross-validation was used to evaluate the learning methods. In each cross-validation trial, the data for one subject was set aside for testing the final model, leaving 18 subjects for training. Prediction accuracy was assessed by the mean squared error (MSE), standard deviation (SD) of the squared error, and Pearson's correlation coefficient ($r$) between the true and predicted outcomes, i.e., change in SRS score. Also, from the predicted outcomes, we computed the relative absolute error (RAE), mean absolute percentage error (MAPE), and SD of the absolute percentage error of the predicted post-treatment SRS scores, the clinical measure of interest. RAE measures error relative to the simple prediction of the average score, while MAPE measures error relative to the true score.

Table 1: Accuracy of predicted treatment response for each learning algorithm.

| Algorithm | MSE ± SD | $p_{MSE}$ | $r$ | $p_r$ | RAE | $p_{RAE}$ | MAPE ± SD | $p_{MAPE}$ |
|---|---|---|---|---|---|---|---|---|
| RF | 0.82 ± 0.96 | 0.019 | 0.39 | 0.038 | 0.63 | 0.044 | 0.24 ± 0.26 | 0.043 |
| SVM | 0.75 ± 0.93 | 0.037 | 0.46 | 0.040 | 0.60 | 0.051 | 0.22 ± 0.22 | 0.051 |
| RF-RF | 0.69 ± 0.80 | 0.024 | 0.54 | 0.023 | 0.56 | 0.026 | 0.21 ± 0.25 | 0.030 |
| RF-B | 0.57 ± 0.67 | 0.012 | 0.68 | 0.006 | 0.50 | 0.012 | 0.20 ± 0.23 | 0.025 |
| RF-BS | 0.40 ± 0.45 | 0.001 | 0.80 | 0.001 | 0.44 | 0.005 | 0.17 ± 0.19 | 0.013 |
| RF-BS-BC | 0.29 ± 0.43 | 0.001 | 0.83 | 0.001 | 0.35 | 0.001 | 0.13 ± 0.15 | 0.001 |

Significance of the observed accuracy measures was assessed using permutation tests. The cross-validation procedure was run 1000 times with the outputs randomly permuted across training samples. Each p-value was computed as the proportion of runs that resulted in values as extreme as the observed statistic.

### 3.3 Results and Discussion

Prediction accuracy for each learning method is listed in Table 1. Plots depicting the true vs. predicted outcomes are shown in Fig. 2. Permutation test results suggest every algorithm learned a regression function which exploited the relationship between the fMRI data and treatment outcomes (although for SVM, post-treatment score errors were not significant at the 0.05 level). The standard random forest and support vector machine algorithms performed the worst. Following random forest variable selection with bagging was more accurate than with random forest. This may be explained by the decreased strength of the individual trees having a greater effect than the decreased variance from reducing the correlation between trees. Prediction accuracy further increased when stepwise variable refinement was included, as less relevant or noisy inputs are filtered out. Finally, the proposed full pipeline resulted in the highest prediction accuracy; including bias correction further reduced the errors by over 20%.

The results of variable selection using the proposed pipeline are shown in Fig. 3. Initial inputs were voxels in brain regions corresponding to social motivation (Fig. 3a). The random forest variable importance ranking step reduced the original 267 voxels to an average of 13 candidate predictors across cross-validation trials (Fig. 3b). From these candidate variables, the stepwise variable refinement stage built a simpler model by removing redundant or noisy voxels, keeping only the voxels that, when combined, significantly reduced prediction error (Fig. 3c). This helps explain the scattered final input voxels, as neighboring voxels likely contain redundant information. An average of 5 voxels were selected as inputs for the final prediction models across cross-validation runs. Note voxel colors indicate frequency of inclusion as a candidate predictor (Fig. 3b) or in the final prediction model (Fig. 3c) across cross-validation trials. The results indicate that voxels in the orbitofrontal cortex and amygdala are the most informative in predicting patient response to PRT. Note that this does not imply that activity in other voxels has no influence on response to treatment.

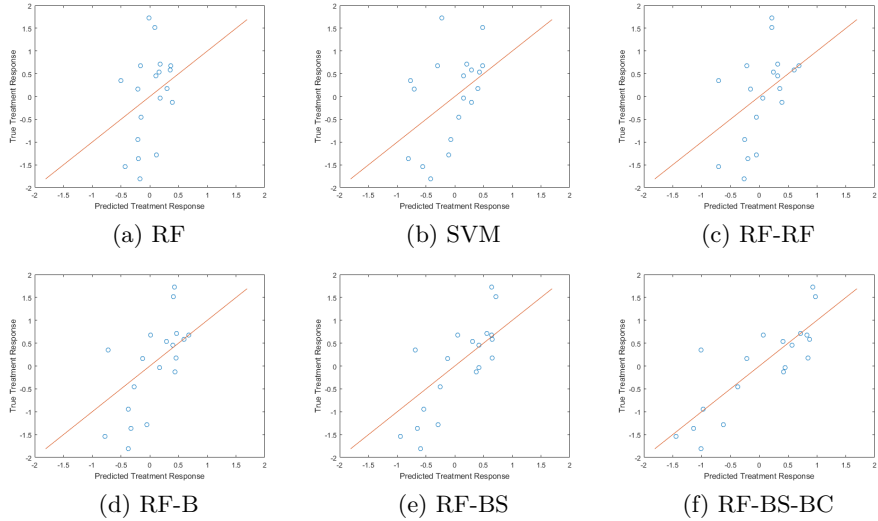

Fig. 2: Plots of the true vs. predicted treatment outcome for each learning method. The red line is a visual guide to show where perfect predictions would lie.

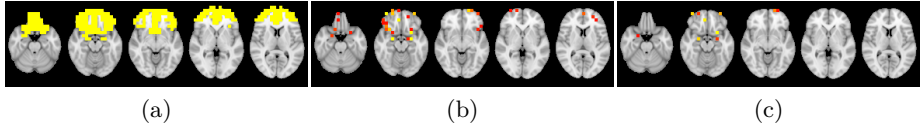

Fig. 3: Variable selection results. Voxels included as (a) initial inputs, (b) candidate predictors using random forest variable ranking, and (c) inputs in the final prediction models using the proposed pipeline. Voxel colors in (b) and (c) denote frequency of selection across cross-validation runs, with red to yellow indicating low to high frequency.

## 4 Conclusions

We presented a learning pipeline based on random forest regression and tree bagging to predict autism therapy response from baseline visual task-based fMRI. The proposed variable selection and bias correction process led to improved prediction accuracy compared to several other methods.

Future work will explore other possible biomarkers for prediction, such as activation in other brain regions or functional connectivity. Also, as more patient data is acquired, the learning approach will be applied to larger datasets to demonstrate generalizability. The ability to predict response to a variety of autism therapies will help inform clinical decisions and personalize treatment.